\documentclass[]{article}

\usepackage[titletoc,toc,title]{appendix}
\usepackage{graphicx}
\usepackage{dcolumn}
\usepackage{bm}

\usepackage{subfig}
\usepackage{caption}
\usepackage{epstopdf}
\usepackage[numbers,sort]{natbib}
\usepackage{graphics}
\usepackage{latexsym}
\usepackage{amssymb}
\usepackage{amsmath}
\usepackage{mathrsfs}
\usepackage{calrsfs} 
\usepackage{rotating}
\usepackage{authblk}
\usepackage{xfrac}
\usepackage{color}

\newcommand{\ie}{{\it i.e.}}

\newcommand*\samethanks[1][\value{footnote}]{\footnotemark[#1]}

\graphicspath{{./pics/}{./picsprs/}}

\date{\today}

\title{Stability and folds in an elastocapillary system} 

\author{Amir Akbari\thanks{Department of Chemical Engineering, McGill University, Montreal, Quebec H3A 0C5} \and Reghan J. Hill\samethanks \and Theo G.M. van de Ven\thanks{Department of Chemistry, McGill University, Montreal, Quebec H3A 2A7}}

\begin{document}
\maketitle

\begin{abstract}
We examine the equilibrium and stability of an elastocapillary system to model drying-induced structural failures. The model comprises a circular elastic membrane with a hole at the center that is deformed by the capillary pressure of simply connected and doubly connected menisci. Using variational and spectral methods, stability is related to the slope of equilibrium branches in the liquid content versus pressure diagram for the constrained and unconstrained problems. The second-variation spectra are separately determined for the membrane and meniscus, showing that the membrane out-of-plane spectrum and the in-plane spectrum at large elatocapillary numbers are both positive, so that only meniscus perturbations can cause instability. At small elastocapillary numbers, the in-plane spectrum has a negative eigenvalue, inducing wrinkling instabilities in thin membranes. In contrast, the smallest eigenvalue of the meniscus spectrum always changes sign at a pressure turning point where stability exchange occurs in the unconstrained problem. We also examine configurations in which the meniscus and membrane are individually stable, while the elastocapillary system as a whole is not; this emphasizes the connection between stability and the coupling of elastic and capillary forces.
\end{abstract}


\section{Introduction} \label{sec:introduction}

Elastic deformations induced by capillary forces have been identified as leading causes of pattern collapse in miniature electronic devices and sensors \citep{chandra2009capillary, chini2010understanding}. These microstructures are more prone to collapse when miniaturized because adhesion and capillary forces become comparable to elastic forces when there is a high surface area to volume ratio \citep{chandra2010stability}. Capillary driven collapse poses major problems for micro-fabrication techniques that are based on wet etching. In particular, microelectronic systems are commonly fabricated through wet lithography where structures often experience permanent deformation and stiction upon drying, significantly limiting the design and operating conditions \citep{roman2010elasto}. 

Elastocapillary systems have been extensively studied over the past two decades \citep{mastrangelo1993mechanical, bico2004adhesion, pokroy2009self, duprat2012wetting, giomi2012minimal}. Aggregation, coalescence, and self-assembly of filaments and flexible fibres \citep{cohen2003kinks, kim2006capillary, boudaoud2007elastocapillary, pokroy2009self}, failure of microelectronic devices \citep{mastrangelo1993mechanicalb, raccurt2004influence}, and capillary wrinkling of elastic membranes \citep{huang2007capillary, vella2010capillary} are applications where system configurations and structures are determined by elastic-capillary force interactions. The foregoing studies are mostly concerned with systems where equilibrium configurations are always stable (or assumed to be). However, mechanical stability is central to applications in which preventing structural failure upon drying is crucial. 

Recently, a few studies have focused on the mechanical stability of elastocapillary systems. \citet{giomi2012minimal} examined the equilibrium and stability of minimal surfaces spanning deformable frames. Subjecting a circular frame to spatial perturbations, they approximated instability modes and the critical elastocapillary numbers corresponding to the primary and secondary buckling of the frame into elliptical and twisted structures. \citet{taroni2012multiple} identified multiple equilibria in an elastocapillary system related to the aggregation of paint-brush bristles where the stable solutions for a given liquid content were determined through a temporal stability and dynamic analysis. 

\citet{mastrangelo1993mechanical} took a different approach to determine stability in elastocapillary systems. Their approach hinges on a pervasive theory, known as catastrophe theory \citep{arnol1992catastrophe} in nonlinear dynamics, stating that stability exchanges only occur at folds and branch points on equilibrium branches \citep{seydel2009practical}. While this has not been generally proved for all mechanical systems, the idea has been extensively examined for purely capillary \citep{myshkis1987low,vogel1989stability,slobozhanin1997bifurcation,akbari2014catenoid,akbari2014bridge} and purely elastic \citep{thompson1984elastic} problems. In this context, \citet{maddocks1987stability} established a theory for systems where equilibria are described by a continuous functional of a single function with prescribed boundary conditions (from a Hilbert space). This theory relates the stability of constrained and unconstrained variational problems to the shape of equilibrium branches with no branch point where stability exchanges occur only at simple folds.

Determining the stability of an elastic structure deformed by the Laplace pressure or contact line force of a meniscus is more challenging than determining its equilibria. Elastic and capillary parts for equilibrium states can be decoupled and determined separately by imposing the proper boundary conditions where the meniscus and structure meet. However, the stability of elastic and capillary parts alone is not sufficient to deduce the stability of elastocapillary systems for which the control-parameter role is particularly important. 

In the present work, we study an elastocapillary problem where conformations are controlled by the liquid content, similarly to \citet{kwon2008equilibrium}. This is analogous to the problem of disconnected free surfaces considered by \citet{slobozhanin1983stability}. Complexities in elastocapillary problems arise because meniscus perturbations are neither pressure controlled nor volume controlled, as in purely capillary problems \citep{lowry1995capillary, akbari2014bridge}. Moreover, elastic structures and menisci in many practical elastocapillary systems have free boundaries \citep{bico2004adhesion, roman2010elasto, taroni2012multiple, duprat2012wetting}, which considerably complicate the stability analysis. \citet{vogel2000sufficient} highlights two major difficulties for analyzing the quadratic forms arising from the second variation of systems with free boundaries: (i) The function space of perturbations is not necessarily a symmetric Hilbert space $\mathscr{H}^{0}$. Instead, the quadratic forms are naturally expressed in $\mathscr{H}^{1}$, and an additional analysis is required to link the arising operators to the corresponding operators in a symmetric $\mathscr{H}^{0}$ space\footnote{Note that $\mathscr{H}^{k}=W^{k,2}$ denotes Sobolev spaces equipped with the Euclidean norm  \citep{adams2003sobolev}. Moreover, throughout the paper, `symmetric space' refers to a space in which all bilinear forms are symmetric.}. (ii) Perturbed surfaces resulting from normal variations of an equilibrium surface are not generally guaranteed to satisfy the boundary conditions at the free boundaries.   

In this paper, we examine the elastocapillary system shown in Fig.~\ref{fig:figure1} and relate stability to the slope of equilibrium branches in pressure versus volume diagrams, similarly to \citet{maddocks1987stability}. This system is a model for drying-induced structural failures arising in practical applications, such as the stiction of micro-machined sensors and collapse of wood fibres upon drying. It comprises a circular elastic membrane, with a hole at the center, anchored above a rigid plate, trapping a prescribed volume of liquid. We examine membrane deformations caused by a meniscus at the hole as the liquid is slowly removed. Our approach is variational, so that linear stability is determined by the sign of the second variation. We demonstrate that there are configurations in which the meniscus and membrane are individually stable, while the elastocapillary system as a whole is not. This emphasizes the significance of instabilities arising from the coupling of elastic and capillary forces. This result can be interpreted as the equivalent of the Weierstrass--Erdmann condition \citep{gelfand2000calculus} for the second variation, and it is relevant to applications where extrema are represented by non-smooth functions, such as for elastocapillary systems, threshold phenomena \citep{clarke1990optimization}, and data visualization \citep{mumford1989optimal}.

\section{Formulation} \label{sec:formulation}
\begin{figure} 
\centering
\includegraphics[width=\linewidth]{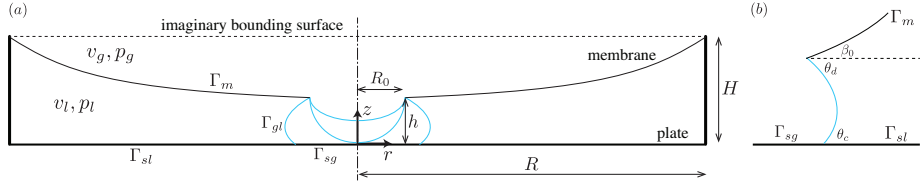}
 \caption{Elastocapillary model; (a) schematic showing simply connected meniscus (top), doubly connected meniscus (bottom), transition from simply to doubly connected meniscus (middle), and (b) contact angles.}
\label{fig:figure1}
 \end{figure}

We consider an elastocapillary model comprising a circular elastic membrane with a hole at the center supported on the sidewall of a cylindrical cavity with rigid walls, trapping a liquid volume $v_{l}$ below the membrane and air volume $v_{g}$ between the bounding surface (dashed line in Fig.~\ref{fig:figure1}(a)) and membrane, as shown in Fig.~\ref{fig:figure1}. The cavity is open to the atmosphere from the top. A meniscus forms at the hole as the liquid is removed, resulting in a difference between the liquid pressure $p_{l}$ and atmospheric pressure $p_{g}$, which causes the membrane to deform. Here, the membrane radius $R$, hole radius $R_{0}$, and cylinder hight $H$ are the model length scales that control the interplay between elastic and capillary forces. To determine the equilibria at a given $v_{l}$, we consider an imaginary bounding surface that covers the cavity from the top. The system is completely isolated from the surrounding by the bounding surface and cylinder walls. The meniscus is initially a bubble, which can bridge the gap upon contact with the plate at the bottom of the cylinder, forming a free contact line with the plate. Assuming that all the dimensions are small compared to the capillary length, the gravity force is neglected. The membrane and meniscus are assumed axisymmetric in equilibrium and perturbed configurations.  

\subsection{Variational principle} \label{sec:variationalprinciple}
Following the development of \citet{neumann2012applied}, we apply a variational principle to determine stable equilibria for the system depicted in Fig.~\ref{fig:figure1}. Noting that the liquid volume is the control parameter in drying, the total internal energy is to be minimized subject to $v_{l}=\mbox{const.}$, maintaining fixed total entropy and mass. Here, the grand canonical potential is a suitable free-energy representation because it restricts the minimization to states that are already in thermal (constant uniform temperature) and chemical (constant uniform chemical potential) equilibrium. The grand canonical potentials for bulk phases and interfaces, respectively, are \citep{neumann2012applied}
\begin{equation}
	\omega^{(v)}=u^{(v)}-Ts^{(v)}-\mu\rho^{(v)}=-p_{v},	\qquad	v=l,g,
    	 \label{eqn:eq1}
\end{equation}
\begin{equation}
	\omega^{(a)}=u^{(a)}-Ts^{(a)}-\mu\rho^{(a)}=\gamma_{a},	\qquad	a=gl,sl,sg
    	 \label{eqn:eq2}
\end{equation}
with $\gamma$, $\omega$, $u$, $s$, and $\rho$ the surface tension, specific grand canonical potential, specific internal energy, specific entropy, and density of the respective phase. The superscripts $(v)$ and $(a)$ denote volume density and area density for bulk phases and interfaces. Note that the temperature $T$ and chemical potential $\mu$ can be regarded as the Lagrange multipliers associated with the entropy and mass in the foregoing constrained minimization of the total internal energy. Hence, stable equilibria minimize 
\begin{equation}
	E_{t}=\omega^{(g)}v_{g}+\omega^{(l)}v_{l}+\omega^{(gl)}\Gamma_{gl}+\omega^{(sl)}\Gamma_{sl}+\omega^{(sg)}\Gamma_{sg}+\Omega^{(m)},
    	 \label{eqn:eq3}
\end{equation}
where $\Gamma_{ij}$ are interfacial surface areas. Here, the membrane strain energy $\Omega^{(m)}$ is separately incorporated into the total energy $E_{t}$ to account for variable and anisotropic stresses. Neglecting the bending energy, we only consider the stretching part of the elastic strain energy in von K\'arm\'an's theory for moderately large deflections 
\begin{equation}
	\Omega^{(m)}=\frac{1}{2}\int_{\Gamma_{m0}} (N_{rr}\varepsilon_{rr}+N_{tt}\varepsilon_{tt}) \mbox{d}A,
    	 \label{eqn:eq4}
\end{equation}
where $\Gamma_{m0}$, $N_{ii}$, and $\varepsilon_{ii}$ are the membrane in the referential configuration (undeflected state), axial forces, and nonlinear strains \citep{timoshenko1959theory}. Substituting Eqs.~(\ref{eqn:eq1}), (\ref{eqn:eq2}), and (\ref{eqn:eq4}) into Eq.~(\ref{eqn:eq3}) and omitting additive and multiplicative constants that do not affect the minimization,
\begin{equation}
	E_{t}[r,u,w,P]=U[r,u,w]-PJ[r,u,w], \quad	E_{t}:L^{2}\times L^{2}\times L^{2}\times\mathbb{R} \rightarrow\mathbb{R}
    	 \label{eqn:eq5}
\end{equation}
is the functional to be minimized subject to $v_{l}=\mbox{const}.$, where
\begin{equation}
	U[r,u,w]=\int_{z_{0}}^{h} F(z,r,r')\mbox{d}z+\int_{R_{00}}^{R} G(r_{p},u,w,u',w')\mbox{d}r_{p}+\frac{R_{1}^{2}}{2}(\gamma_{sg}-\gamma_{sl}),
    	 \label{eqn:eq6}
\end{equation}
\begin{equation}
	J[r,u,w]=\int_{z_{0}}^{h} K(z,r,r')\mbox{d}z+\int_{R_{00}}^{R} M(r_{p},u,w,u',w')\mbox{d}r_{p}+\frac{R_{0}^{2}h}{2}
    	 \label{eqn:eq7}
\end{equation}
with integrands 
\begin{equation}
	F(z,r,r')=\gamma_{gl}r\sqrt{1+r'^{2}},
    	 \label{eqn:eq8}
\end{equation}
\begin{equation}
	K(z,r,r')=-\frac{r^{2}}{2},
    	 \label{eqn:eq9}
\end{equation}
\begin{equation}
	G(r_{p},u,w,u',w')=\frac{C}{2}\left(u'^{2}+u'w'^{2}+\frac{2\nu uu'}{r_{p}}+\frac{\nu uw'^{2}}{r_{p}}+\frac{u^{2}}{r_{p}^{2}}+\frac{w'^{4}}{4}  \right)r_{p},
    	 \label{eqn:eq10}
\end{equation}
\begin{equation}
	M(r_{p},u,w,u',w')=uH+(r_{p}+u)H+(r_{p}+u)(H+w)u'.
    	 \label{eqn:eq11}
\end{equation}
Note that $C$, $\nu$, $u$, $w$, $R_{1}$, and $R_{00}$ are the membrane axial rigidity, Poisson ratio, membrane in-plane displacement\footnote{Not to be confused with the specific internal energy in Eqs.~(\ref{eqn:eq1}) and (\ref{eqn:eq2}).}, membrane deflection, radius of the meniscus contact line with the plate, and hole radius in the referential configuration. Here, primes denote derivatives with respect to the function argument, $P=p_{l}-p_{g}$ can be regarded as the Lagrange multiplier associated with the constant $v_{l}$ constraint, and the menisci are represented by $r(z)$. The membrane deformations are represented by $u(r_{p})$ and $w(r_{p})$, where $r_{p}$ is the radial coordinate in the referential configuration. When the meniscus is a bubble (simply connected), the last term in Eq.~(\ref{eqn:eq6}) is zero and $z_{0}=\ell$, where the meniscus intersects the symmetry axis. When the meniscus is a bridge (doubly connected), $z_{0}=0$, where the free contact line rests on the plate. 

\subsection{Equilibrium from first variation} \label{sec:equilibrium}
To construct the increment of $E_{t}$ in Eq.~(\ref{eqn:eq5}) with respect to axisymmetric perturbations, perturbed states are represented by 

\begin{equation}
	z(\hat{z})=\hat{z}+\eta_{1}(\hat{z})\varepsilon+\eta_{2}(\hat{z})\varepsilon^{2}, \qquad \eta_{1},\eta_{2}:[\hat{z}_{0},\hat{h}]\rightarrow\mathbb{R},
    	 \label{eqn:eq12}
\end{equation}
\begin{equation}
	r(\hat{z})=\hat{r}(\hat{z})+\xi_{1}(\hat{z})\varepsilon+\xi_{2}(\hat{z})\varepsilon^{2}, \qquad \hat{r},\xi_{1},\xi_{2}:[\hat{z}_{0},\hat{h}]\rightarrow\mathbb{R},
    	 \label{eqn:eq13}
\end{equation}
\begin{equation}
	u(r_{p})=\hat{u}(r_{p})+\phi_{1}(r_{p})\varepsilon+\phi_{2}(r_{p})\varepsilon^{2}, \qquad \hat{u},\phi_{1},\phi_{2}:[R_{00},R]\rightarrow\mathbb{R},
    	 \label{eqn:eq14}
\end{equation}
\begin{equation}
	w(r_{p})=\hat{w}(r_{p})+\psi_{1}(r_{p})\varepsilon+\psi_{2}(r_{p})\varepsilon^{2}, \qquad \hat{w},\psi_{1},\psi_{2}:[R_{00},R]\rightarrow\mathbb{R},
    	 \label{eqn:eq15}
\end{equation}
accounting for the linear and nonlinear parts of the increment when the meniscus is a bridge. Note that the form of the functionals in Eqs.~(\ref{eqn:eq6}) and ~(\ref{eqn:eq7}) demands $\hat{r}', \hat{u}', \hat{w}'$ to be continuous, so $\hat{r}, \hat{u}, \hat{w}\in L^{2}\cap C^{1}$. Here, equilibrium and perturbed states are denoted by hatted and unhatted variables, respectively. Equation~(\ref{eqn:eq12}) is particularly important, because it admits perturbations that can displace the position of the bubble apex and hole edge along the $z$-axis. We impose a simply supported boundary condition for the membrane at $r_{p}=R$ where $u,w=0$ \citep{timoshenko1959theory}, resulting in 
\begin{equation}
	\phi_{1}(R)=\phi_{2}(R)=\psi_{1}(R)=\psi_{2}(R)=0.
    	 \label{eqn:eq16}
\end{equation}
The meniscus is assumed to be pinned to the hole edge at $r_{p}=R_{00}$, where $r(\hat{h})|_{\Gamma_{gl}}=r(R_{00})|_{\Gamma_{m}}$ and $z(\hat{h})|_{\Gamma_{gl}}=z(R_{00})|_{\Gamma_{m}}$, furnishing 
\begin{equation}
	\hat{R_{0}}=R_{00}+\hat{u}(R_{00}),\quad \xi_{1}(\hat{h})=\phi_{1}(R_{00}), \quad \xi_{2}(\hat{h})=\phi_{2}(R_{00}),
    	 \label{eqn:eq17}
\end{equation}
\begin{equation}
	\hat{h}=H+\hat{w}(R_{00}),\quad \eta_{1}(\hat{h})=\psi_{1}(R_{00}), \quad \eta_{2}(\hat{h})=\psi_{2}(R_{00}).
    	 \label{eqn:eq18}
\end{equation}
Moreover, the meridian curve intersects the symmetry axis at $\hat{z}=\hat{\ell}$, where it can only move vertically when the meniscus is a bubble, so
\begin{equation}
	\xi_{1}(\hat{\ell})=\xi_{2}(\hat{\ell})=0, \quad \eta_{1}(\hat{\ell}),\eta_{2}(\hat{\ell})=\mbox{finite},
    	 \label{eqn:eq19}
\end{equation}
whereas the contact line with the plate can only move horizontally at $\hat{z}=0$ when the meniscus is a bridge, so
\begin{equation}
	\xi_{1}(0),\xi_{2}(0)=\mbox{finite}, \quad \eta_{1}(0)=\eta_{2}(0)=0.
    	 \label{eqn:eq20}
\end{equation}

Since the domain of $r(z)$ is variable in Eqs.~(\ref{eqn:eq6}) and (\ref{eqn:eq7}), the functional variations are properly represented with respect to the barred component of $\xi$ (see \citet{gelfand2000calculus} for details)
\begin{equation}
	\xi_{1}=\bar{\xi}_{1}+\eta_{1}\hat{r}',
    	 \label{eqn:eq21}
\end{equation}
\begin{equation}
	\xi_{2}=\bar{\xi}_{2}+\eta_{2}\hat{r}'+\eta_{1}\xi'_{1}+\frac{1}{2}\eta_{1}^2\hat{r}''.
    	 \label{eqn:eq22}
\end{equation}
Substituting Eqs.~(\ref{eqn:eq12})-(\ref{eqn:eq15}) into Eq.~(\ref{eqn:eq5}), the first variation of $E_{t}$ with respect to an equilibrium state is
\begin{multline}
	\frac{\delta E_{t}}{\varepsilon}=\left<U'_{(\hat{r})}-PJ'_{(\hat{r})},\bar{\xi}_{1}\right>+\left<U'_{(\hat{u})}-PJ'_{(\hat{u})},\phi_{1}\right>+\left<U'_{(\hat{w})}-PJ'_{(\hat{w})},\psi_{1}\right>\\
	+\left[ F_{r'}|_{\hat{h}}-P\hat{R}_{0}\hat{h}-G_{u'}|_{R_{00}}+PM_{u'}|_{R_{00}} 	\right]\xi_{1}(\hat{h})\\
	+\left[ F|_{\hat{h}}-\hat{r}'F_{r'}|_{\hat{h}}-G_{w'}|_{R_{00}} \right]\eta_{1}(\hat{h})+\left[ \hat{R}_{1}(\gamma_{sg}-\gamma_{sl})-F_{r'}|_{0}\right]\xi_{1}(0),
    	 \label{eqn:eq23}
\end{multline}
where all the functional integrands are evaluated at the equilibrium. Here, primes operating on functionals denote the first Fr\'echet derivative \citep{bobylov1999geometrical} with respect to the function in the subscript, and $\left<\cdot,\cdot\right>$ is the inner product over the domain of the respective function. The last term in Eq.~(\ref{eqn:eq23}) must be replaced with $\left[\hat{r}'F_{r'}-F\right]_{\hat{\ell}}\eta_{1}(\hat{\ell})$ when the meniscus is a bubble. Equilibria are the stationary points of the total energy where $\delta E_{t}=0$ for arbitrary $\bar{\xi}_{1}$, $\phi_{1}$, and $\psi_{1}$, requiring 
 \begin{equation}
	U'_{(\hat{r})}-PJ'_{(\hat{r})}=F_{r}-PK_{r}-\frac{\mbox{d}}{\mbox{d}\hat{z}}\left(F_{r'}-PK_{r'} \right)=0,
    	 \label{eqn:eq24}
\end{equation}
\begin{equation}
	U'_{(\hat{u})}-PJ'_{(\hat{u})}=G_{u}-PM_{u}-\frac{\mbox{d}}{\mbox{d}r_{p}}\left(G_{u'}-PM_{u'} \right)=0,
    	 \label{eqn:eq25}
\end{equation}
\begin{equation}
	U'_{(\hat{w})}-PJ'_{(\hat{w})}=G_{w}-PM_{w}-\frac{\mbox{d}}{\mbox{d}r_{p}}\left(G_{w'}-PM_{w'} \right)=0,
    	 \label{eqn:eq26}
\end{equation}
with each boundary term in square brackets equal to zero. Substituting the integrands from Eqs.~(\ref{eqn:eq8})-(\ref{eqn:eq11}) furnishes 
 \begin{equation}
	\frac{\hat{r}''}{(1+\hat{r}'^2)^{3/2}}-\frac{1}{\hat{r}(1+\hat{r}'^2)^{1/2}}=\frac{P}{\gamma_{gl},}
    	 \label{eqn:eq27}
\end{equation}
\begin{equation}
	r_{p}\hat{N}'_{rr}+\hat{N}_{rr}-\hat{N}_{tt}-P(r_{p}+\hat{u})\hat{w}'=0,
	\label{eqn:eq28}
\end{equation}
\begin{equation}
	(\hat{N}_{rr}\hat{w}'r_{p})'+P(r_{p}+\hat{u})(1+\hat{u}')=0
    	 \label{eqn:eq29}
\end{equation}
with boundary conditions
 \begin{equation}
	\gamma_{gl}\hat{R}_{0}\cos\theta_{d}=-\hat{N}_{rr}(R_{00})R_{00}\quad \mbox{at} \quad \hat{z}=\hat{h},
    	 \label{eqn:eq30}
\end{equation}
\begin{equation}
	\gamma_{gl}\hat{R}_{0}\sin\theta_{d}=\hat{N}_{rr}(R_{00})\hat{w}'(R_{00})R_{00} \quad \mbox{at} \quad \hat{z}=\hat{h},
	\label{eqn:eq31}
\end{equation}
\begin{equation}
	\gamma_{sg}-\gamma_{sl}=\gamma_{gl}\cos\theta_c \quad \mbox{at} \quad \hat{z}=0.
    	 \label{eqn:eq32}
\end{equation}
Here, $\theta_{c}$ and $\theta_{d}$ are the contact and dihedral angles that the interface $\Gamma_{gl}$ forms with the plate and membrane, respectively. The last boundary condition only holds when the meniscus is a bridge, while the boundary term associated with $\eta_{1}(\hat{\ell})$ when the meniscus is a bubble is always zero with $\hat{r}'(\hat{\ell})\to\infty$. Equation~(\ref{eqn:eq27}) is the Young-Laplace equation, and Eqs.~(\ref{eqn:eq28}) and (\ref{eqn:eq29}) are the in- and out-of-plane equations of equilibrium for membranes in von K\'arm\'an's theory with the capillary pressure acting normal to the neutral plane. Note that Eqs.~(\ref{eqn:eq30}) and (\ref{eqn:eq31}) demand $\hat{w}'(R_{00})\hat{r}'(\hat{h})=1$, implying $\hat{w}'\to\infty$ as $\theta_{d}\to\pi/2$, contradicting the assumptions of von K\'arm\'an's theory \citep{timoshenko1959theory}. To resolve this issue, we undertake a scaling analysis in section \ref{sec:scaling} to simplify Eqs.~(\ref{eqn:eq30}) and (\ref{eqn:eq31}).

\begin{figure} 
\centering
\includegraphics[width=0.8\linewidth]{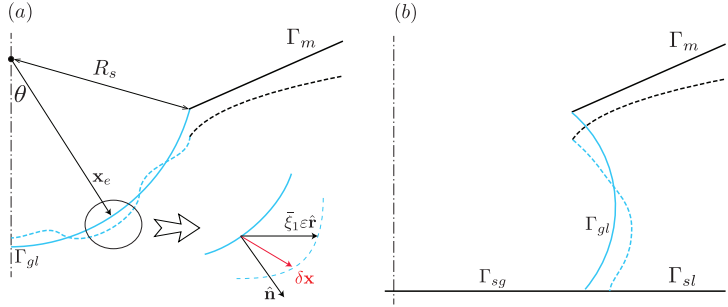}
 \caption{Schematic of perturbations to (a) simply connected (bubble), and (b) doubly connected (bridge) menisci.}
\label{fig:figure2}
 \end{figure}
 
\subsection{Stability from second variation} \label{sec:stability}
The Jacobian matrices of the functional integrants in Eqs.~(\ref{eqn:eq6}) and (\ref{eqn:eq7}) are not symmetric. Since analyzing the second variation for functionals of multiple functions with non-symmetric Jacobians is intractable \citep{gelfand2000calculus}, we simplify the problem by neglecting in-plane variations at the hole edge, prescribing $\phi_{i}(R_{00})=0$. Accordingly, the second variation is
\begin{multline}
	\frac{2\delta^{2} E_{t}}{\varepsilon^2}=\int_{\hat{z}_{0}}^{\hat{h}} \left(\mathcal{P}^{(\hat{r})}\bar{\xi}_{1}'^{2}+\mathcal{Q}^{(\hat{r})}\bar{\xi}_{1}^{2}\right)\mbox{d}\hat{z}+\int_{R_{00}}^{R} \left(\mathcal{P}^{(\hat{u})}\phi_{1}'^{2}+\mathcal{Q}^{(\hat{u})}\phi_{1}^{2}\right)\mbox{d}r_{p}\\
	+\int_{R_{00}}^{R} \left(\mathcal{P}^{(\hat{w})}\psi_{1}'^{2}+\mathcal{Q}^{(\hat{w})}\psi_{1}^{2}\right)\mbox{d}r_{p}+[\mathcal{A}\bar{\xi}_{1}^{2}]_{\hat{h}}-[\mathcal{A}\bar{\xi}_{1}^{2}]_{\hat{z}_{0}},
    	 \label{eqn:eq33}
\end{multline}
where $\mathcal{F}=F-PK$, $\mathcal{G}=G-PM$, and
\begin{equation}
	\mathcal{P}^{(\hat{r})}=\mathcal{F}_{r'r'},\quad \mathcal{Q}^{(\hat{r})}=\mathcal{F}_{rr}-\frac{\mbox{d}}{\mbox{d}\hat{z}} \mathcal{F}_{rr'},
    	 \label{eqn:eq34}
\end{equation}
\begin{equation}
	\mathcal{P}^{(\hat{u})}=\mathcal{G}_{u'u'},\quad \mathcal{Q}^{(\hat{u})}=\mathcal{G}_{uu}-\frac{\mbox{d}}{\mbox{d}r_{p}} \mathcal{G}_{uu'},
    	 \label{eqn:eq35}
\end{equation}
\begin{equation}
	\mathcal{P}^{(\hat{w})}=\mathcal{G}_{w'w'},\quad \mathcal{Q}^{(\hat{w})}=\mathcal{G}_{ww}-\frac{\mbox{d}}{\mbox{d}r_{p}} \mathcal{G}_{ww'},
    	 \label{eqn:eq36}
\end{equation}
\begin{equation}
	\mathcal{A}=\mathcal{F}_{rr'}-\frac{\mathcal{F}_{r}}{\hat{r}'}.
    	 \label{eqn:eq37}
\end{equation}
Note that the boundary term at $\hat{z}_{0}$ in Eq.~(\ref{eqn:eq33}) arises only when the meniscus is a bubble, so $\delta^{2} E_{t}$ has only one boundary term at $\hat{h}$ when the meniscus is a bridge. Furthermore, when the meniscus is a bubble, $\bar{\xi}_{1}$ is not bounded due to the axial symmetry condition at $\hat{z}_{0}=\hat{\ell}$ (see Appendix~\ref{sec:appendixa}), and it is unsuitable for representing quadratic forms. To resolve this issue, using the mapping $(\hat{z},\bar{\xi}_{1})\rightarrow(y,N)$, the first integral and boundary terms in Eq.~(\ref{eqn:eq33}) are represented with respect to the normal variation $N(y)$, where $y=\cos \theta$ (see Fig.~\ref{fig:figure2}(a) and Appendix~\ref{sec:appendixa}). Equation~(\ref{eqn:eq33}) then reduces to
\begin{multline}
	Q=\gamma_{gl}\int_{y_{0}}^{1} \left[(1-y^{2})N'^{2}+2yNN'-N^{2}\right]\mbox{d}y+\gamma_{gl}\frac{N^{2}(y_{0})}{y_{0}}-\gamma_{gl}N^{2}(1)\\
	+\int_{R_{00}}^{R} \left[r_{p}C\phi_{1}'^{2}+\left(\frac{C}{r_{p}}+P\hat{w}' \right)\phi_{1}^{2}\right]\mbox{d}r_{p}
	+\int_{R_{00}}^{R} (\hat{N}_{rr}+C\hat{w}'^{2})r_{p}\psi_{1}'^{2}\mbox{d}r_{p},
    	 \label{eqn:eq38}
\end{multline}
\begin{multline}
	Q=\int_{0}^{\hat{h}} \left[\mathcal{P}^{(\hat{r})}\bar{\xi}_{1}'^{2}+\mathcal{Q}^{(\hat{r})}\bar{\xi}_{1}^{2}\right]\mbox{d}\hat{z}+[\mathcal{A}\bar{\xi}_{1}^{2}]_{\hat{h}}\\
	+\int_{R_{00}}^{R} \left[r_{p}C\phi_{1}'^{2}+\left(\frac{C}{r_{p}}+P\hat{w}' \right)\phi_{1}^{2}\right]\mbox{d}r_{p}
	+\int_{R_{00}}^{R} (\hat{N}_{rr}+C\hat{w}'^{2})r_{p}\psi_{1}'^{2}\mbox{d}r_{p},
    	 \label{eqn:eq39}
\end{multline}
for simply connected and doubly connected menisci, respectively. Here, $Q=2\delta^{2} E_{t}/\varepsilon^2$, $y_{0}=\cos\theta_{0}$ corresponding to the polar angle at the hole edge (Fig.~\ref{fig:figure2}(a)), and 
\begin{equation}
	\mathcal{A}=-\frac{\gamma_{gl}\hat{r}\hat{r}''}{\hat{r}'(1+\hat{r}'^{2})^{3/2}},\quad \mathcal{P}^{(\hat{r})}=\frac{\gamma_{gl}\hat{r}}{(1+\hat{r}'^{2})^{3/2}},\quad \mathcal{Q}^{(\hat{r})}=-\frac{\gamma_{gl}}{\hat{r}(1+\hat{r}'^{2})^{1/2}}.
    	 \label{eqn:eq40}
\end{equation}

A necessary (sufficient) condition for $E_{t}$ to have a minimum is that $Q$ be non-negative (strongly positive) for an equilibrium solution \citep{gelfand2000calculus}. Here, strong positivity, referred to as nonlinear stability in the literature \citep{vogel1999non}, must be distinguished from positive-definiteness. According to \citet{vogel1996constrained}, $Q>0$ does not imply a strict local minimum for constrained infinite-dimensional problems. Nevertheless, defining stable equilibria as those for which $Q>0$ holds, \citet{maddocks1987stability} derived sufficient conditions for the positive-definiteness of the second variation. Interestingly, \citet{vogel1996constrained} showed that these conditions are sufficient for Madoccks' functional to have a strict minimum. Therefore, following \citet{maddocks1987stability}, we adopt Madoccks' definition of stability to relate stability to the slope of equilibrium branches. 

Our analysis differs from Madoccks' theory in two respects: (i) The elastocapillary energy $E_{t}$ is a functional of multiple functions, the stationary points of which are represented by non-smooth functions. Moreover, $E_{t}$ has three separate spectra, which complicates the relationship between stability exchanges and equilibrium-branch turning points. (ii) The boundary condition at the hole edge, where the membrane and meniscus meet, is not prescribed. Here, perturbations are arbitrarily finite, posing the same difficulties as discussed by \citet{myshkis1987low} and \citet{vogel2000sufficient} for analyzing the stability of capillary surfaces with free contact lines. 
 
The quadratic forms in Eqs.~(\ref{eqn:eq38}) and (\ref{eqn:eq39}) demand $N,\bar{\xi}_{1},\phi_{1},\psi_{1}\in\mathscr{H}^{1}$, where perturbations satisfy the boundary conditions Eqs.~(\ref{eqn:eq16})-(\ref{eqn:eq20}). Since each bilinear term in Eqs.~(\ref{eqn:eq38}) and (\ref{eqn:eq39}) is bounded, $Q$ can be represented as \citep{akhiezer1993theory,vogel2000sufficient}
\begin{equation}
	Q=\gamma_{gl}\left<\bar{\mathcal{L}}_{(\hat{r})} N,N\right>_{1}+\left<\bar{\mathscr{L}}_{(\hat{u})} \phi_{1},\phi_{1}\right>_{1}+\left<\bar{\mathscr{L}}_{(\hat{w})} \psi_{1},\psi_{1}\right>_{1},
    	 \label{eqn:eq41}
\end{equation}
\begin{equation}
	Q=\left<\bar{\mathscr{L}}_{(\hat{r})} \bar{\xi}_{1},\bar{\xi}_{1}\right>_{1}+\left<\bar{\mathscr{L}}_{(\hat{u})} \phi_{1},\phi_{1}\right>_{1}+\left<\bar{\mathscr{L}}_{(\hat{w})} \psi_{1},\psi_{1}\right>_{1}
    	 \label{eqn:eq42}
\end{equation}
for simply connected and doubly connected menisci, where $\left<\cdot,\cdot\right>_{1}$ is the $\mathscr{H}^{1}$ inner product \citep{adams2003sobolev} and $\bar{\mathcal{L}}_{(\hat{r})},\bar{\mathscr{L}}_{(\hat{r})},\bar{\mathscr{L}}_{(\hat{u})},\bar{\mathscr{L}}_{(\hat{w})}$ are uniquely determined by the respective bilinear terms in Eqs.~(\ref{eqn:eq38}) and (\ref{eqn:eq39}). Because perturbations are arbitrary at the hole edge, the $\mathscr{H}^{1}$ spaces from which perturbations are selected are not symmetric, and, thus, $\bar{\mathcal{L}}_{(\hat{r})},\bar{\mathscr{L}}_{(\hat{r})},\bar{\mathscr{L}}_{(\hat{u})},\bar{\mathscr{L}}_{(\hat{w})}$ are not generally self-adjoint. On the other hand, integrating Eqs.~(\ref{eqn:eq38}) and (\ref{eqn:eq39}) by parts leads to
\begin{equation}
	Q=\gamma_{gl}\left<\mathcal{L}_{(\hat{r})} N,N\right>+\left<\mathscr{L}_{(\hat{u})} \phi_{1},\phi_{1}\right>+\left<\mathscr{L}_{(\hat{w})} \psi_{1},\psi_{1}\right>,
    	 \label{eqn:eq43}
\end{equation}
\begin{equation}
	Q=\left<\mathscr{L}_{(\hat{r})} \bar{\xi}_{1},\bar{\xi}_{1}\right>+\left<\mathscr{L}_{(\hat{u})} \phi_{1},\phi_{1}\right>+\left<\mathscr{L}_{(\hat{w})} \psi_{1},\psi_{1}\right>
    	 \label{eqn:eq44}
\end{equation}
for simply connected and doubly connected menisci, where $\mathscr{L}_{(\hat{y})}\equiv U''_{(\hat{y})}-PJ''_{(\hat{y})}$ with double primes denoting the second Fr\'echet derivative and perturbations satisfying the boundary conditions
\begin{equation}
	N'(y_{0})-N(y_{0})/y_{0}=0,\quad N(1)=\mbox{finite},
    	 \label{eqn:eq45}
\end{equation}
\begin{equation}
	\bar{\xi}'_{1}(0)=0,\quad \bar{\xi}'_{1}(\hat{h})+[\mathcal{A}/\mathcal{P}^{(\hat{r})}]\bar{\xi}_{1}(\hat{h})=0,
    	 \label{eqn:eq46}
\end{equation}
\begin{equation}
	\phi_{1}(R_{00})=0,\quad \phi_{1}(R)=0,
    	 \label{eqn:eq47}
\end{equation}
\begin{equation}
	\psi'_{1}(R_{00})=0,\quad \psi_{1}(R)=0,
    	 \label{eqn:eq48}
\end{equation}
furnishing a symmetric $\mathscr{H}^{0}$ for evaluating $Q$. Here,
\begin{equation}
	\mathcal{L}_{(\hat{r})} N=-\frac{\mbox{d}}{\mbox{d}y}\left[ (1-y^2)\frac{\mbox{d}N}{\mbox{d}y}\right]-2N,
    	 \label{eqn:eq49}
\end{equation}
\begin{equation}
	\mathscr{L}_{(\hat{r})} \bar{\xi}_{1}=-\frac{\mbox{d}}{\mbox{d}\hat{z}}\left[ \mathcal{P}^{(\hat{r})}\frac{\mbox{d}\bar{\xi}_{1}}{\mbox{d}\hat{z}}\right]+\mathcal{Q}^{(\hat{r})}\bar{\xi}_{1},
    	 \label{eqn:eq50}
\end{equation}
\begin{equation}
	\mathscr{L}_{(\hat{u})} \phi_{1}=-\frac{\mbox{d}}{\mbox{d}r_{p}}\left[ r_{p}C\frac{\mbox{d}\phi_{1}}{\mbox{d}r_{p}}\right]+\left( \frac{C}{r_{p}}+P\hat{w}' \right)\phi_{1},
    	 \label{eqn:eq51}
\end{equation}
\begin{equation}
	\mathscr{L}_{(\hat{w})} \psi_{1}=-\frac{\mbox{d}}{\mbox{d}r_{p}}\left[ r_{p}(\hat{N}_{rr}+C\hat{w}'^{2})\frac{\mbox{d}\psi_{1}}{\mbox{d}r_{p}}\right],
    	 \label{eqn:eq52}
\end{equation}
subject to Eqs.~(\ref{eqn:eq45})-(\ref{eqn:eq48}) are regular Sturm-Liouville operators; thus, they are self-adjoint and Fredholm \citep{walter1998ordinary}. Establishing a relationship between the spectrum of the barred operators on $\mathscr{H}^{1}$ and those of unbarred operators on $\mathscr{H}^{0}$ significantly simplifies the analysis, providing a setting to apply the well-studied Sturm-Liouville theory. Lemma 2.5 of \citet{vogel1999non} furnishes this relationship by stating that the barred operators have the same number of negative and non-positive eigenvalues as the corresponding unbarred operators. Moreover, the constant-volume constraint ($v_{l}=\mbox{const}.$) can be written
\begin{equation}
	\left<J'_{(\hat{r})},\bar{\xi}_{1}\right>+\left<J'_{(\hat{u})},\phi_{1}\right>+\left<J'_{(\hat{w})},\psi_{1}\right>=0
	\label{eqn:eq53}
\end{equation}
since $\delta v_{l}=\delta J$. 

\section{Scaling analysis} \label{sec:scaling}
Starting from Eq.~(\ref{eqn:eq28}) and taking $R_{00}\ll R$, we begin by scaling all lengths with $R$. Given $\hat{N}_{rr}/C=\tilde{u}'+\tilde{w}'^{2}/2+\nu\tilde{u}/\tilde{r}_{p}$ and $\hat{N}_{tt}/C=\nu\tilde{u}'+\nu\tilde{w}'^{2}/2+\tilde{u}/\tilde{r}_{p}$, where $\tilde{u}=\hat{u}/R$, $\tilde{w}=\hat{w}/R$, and $\tilde{r}_{p}=r_{p}/R$, we find $\tilde{u}\sim\tilde{w}^{2}$ to balance all the terms in the in-plane equilibrium. Considering the bending and stretching parts of the strain energy in von K\'arm\'an's theory for axisymmetric plates \citep{timoshenko1959theory}
\begin{equation}
	\Omega_{B}=\frac{D}{2}\int_{R_{00}}^{R}\left(w''^2+\frac{w'^2}{r_{p}^2}+\frac{2\nu w'w''}{r_{p}} \right)r_{p}\mbox{d}r_{p},
	\nonumber
\end{equation}
\begin{equation}
	\Omega_{S}=\frac{C}{2}\int_{R_{00}}^{R}\left(u'^{2}+u'w'^{2}+\frac{2\nu uu'}{r_{p}}+\frac{\nu uw'^{2}}{r_{p}}+\frac{u^{2}}{r_{p}^{2}}+\frac{w'^{4}}{4}  \right)r_{p}\mbox{d}r_{p},
	\nonumber
\end{equation}
and noting that $r_{p}\sim R$, $w\sim H$, $C\sim Eb$, and $D\sim Eb^{3}$, where $D$ and $b$ are the bending rigidity and plate thickness, one infers $\Omega_{B}\ll \Omega_{S}$ when $b/H\ll1$ by comparing the energy scales $\Omega_{B}\sim DH^2/R^2$ and $\Omega_{S}\sim CH^4/R^2$. For thin membranes, this justifies neglecting the bending energy compared to the stretching energy. 

Given $N_{rr}\sim C(w/R)^{2}$, $|\tilde{w}|\sim(|Q_{c}|/\kappa N_{C})^{1/3}$ follows from the out-of-plane equilibrium, where $\kappa=R_{00}/R$, $Q_{c}=PR_{00}/\gamma_{gl}$, $N_{C}=C/\gamma_{gl}$ are the scaled hole radius, scaled capillary pressure and elastocapillary number. Noting that $|\tilde{w}|+\kappa\Lambda=\Pi$, the elastocapillary number corresponding to a specific state of the system in Fig.~\ref{fig:figure1} can be estimated as $N_{C}\sim|Q_{c}|/\kappa(\Pi-\kappa\Lambda)^{3}$, where $\Lambda=h/R_{00}$ and $\Pi=H/R$ are the meniscus slenderness and aspect ratio. For example, when the meniscus is a bubble contacting the plate, before bridging the gap, as will be discussed elsewhere, $Q_{c}=-2\sin\theta_{d}$ and $\Lambda=\cot (\theta_{d}/2)$; thus, at $\theta_{d}=\theta_{c}$, which is the critical dihedral angle below which the bubble cannot be stably pinned to the hole edge \citep{myshkis1987low}, we have
\begin{equation}
	N_{C}\sim\frac{2\sin\theta_{c}}{\kappa(\Pi-\kappa\cot \theta_{c}/2)^{3}}.
	\label{eqn:eq54}
\end{equation}

From Eq.~(\ref{eqn:eq30}), $\hat{N}_{rr}(R_{00})\sim\gamma_{gl}$, while $\hat{N}_{rr}\sim C\tilde{w}^2$ in the main body of the membrane. Thus, $\hat{N}_{rr}(R_{00})\ll\hat{N}_{rr}$ when $N_{C}\gg1/\Pi^{2}$, implying that Eqs.~(\ref{eqn:eq30}) and (\ref{eqn:eq31}) can be approximated by a free-edge boundary condition for slender cavities when $N_{C}$ is large.  

\begin{figure} 
\centering
\includegraphics[width=\linewidth]{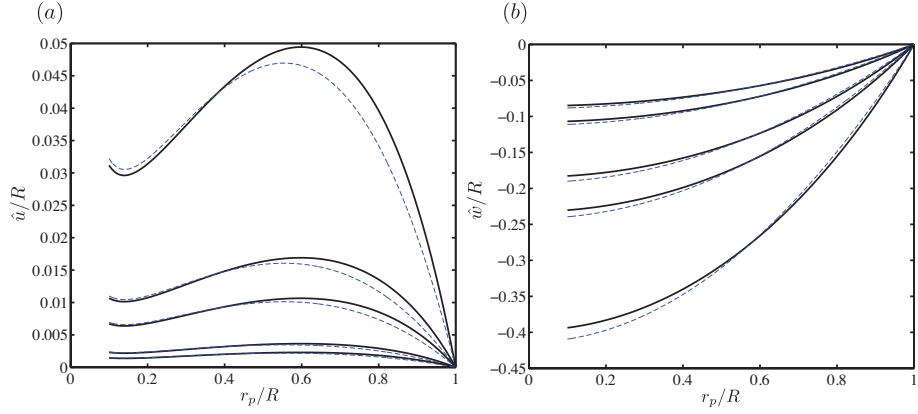}
 \caption{Comparison of the numerically exact solution (solid) and variational approximation (dashed) of (a) in-plane and (b) out-of-plane displacements: $Q_c=-2$, $\kappa=0.1$, $\nu=0.3$, and $N_C=100, 500, 1000, 5000, 10000$ (downward in (a) and upward in (b)).}
\label{fig:figure3}
\end{figure}

\section{Membrane profile} \label{sec:profile}
The stretching part of the strain energy leads to nonlinear equations of equilibrium, which can be solved numerically in most practical problems. Although the case considered in this paper, namely, axisymmetric plate with a hole at the center, has a series solution \citep{timoshenko1959theory}, expressions for the unknown coefficients are cumbersome. Therefore, we apply a variational approximation, as commonly used in the literature \citep{banerjee1981new,mastrangelo1993mechanical}, to construct a general solution for the membrane deflection. 

\begin{figure} 
\centering
\includegraphics[width=\linewidth]{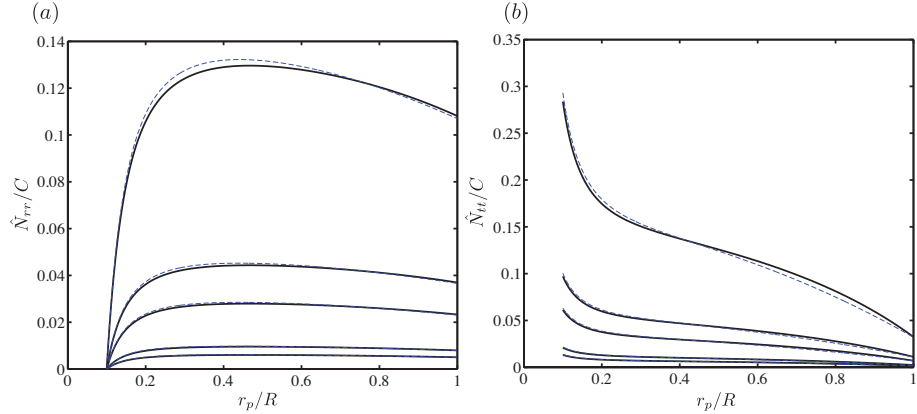}
 \caption{Comparison of the numerically exact solution (solid) and variational approximation (dashed) of axial forces in the (a) radial and (b) tangential directions: $Q_c=-2$, $\kappa=0.1$, $\nu=0.3$, and $N_C=100, 500, 1000, 5000, 10000$ (downward).}
\label{fig:figure4}
\end{figure}

Here, we approximate the boundary condition at $r_{p}=R_{00}$ as a free edge. Noting that the test function 
\begin{equation}
	\tilde{w}=\tilde{w}_{0}(1-\tilde{r}_{p}^{2})
	\label{eqn:eq55}
\end{equation}
is consistent with the boundary conditions and the plate stress distribution under tension at zero deflection \citep{timoshenko1959theory}, $\tilde{w}_{0}=(Q_{c}/N_{C})^{1/3}K_{w}$ is obtained by minimizing the stretching energy, where
\begin{equation}
	K_{w}=\left[ \frac{3[1-\nu+\kappa^{2}(1+\nu)]}{\kappa(1-\kappa^{2})(1-\nu^{2})[7-\nu+\kappa^2(1+\nu)]} \right]^{1/3},
	\label{eqn:eq56}
\end{equation}
which is in agreement with the scaling relation for $\tilde{w}$ derived in section~\ref{sec:scaling}. Equation~(\ref{eqn:eq55}) contrasts with the test function $\tilde{w}=\tilde{w}_{0}(1-\tilde{r}_{p}^{2})^{2}$ that \citet{mastrangelo1993mechanical} used to describe the bending and stretching contribution to the overall deflection of beams. The approximation accuracy hinges on choosing appropriate test functions for bending- and stretching-dominated regimes. Figures~\ref{fig:figure3} and \ref{fig:figure4} demonstrate a reasonable agreement between the variational approximation and numerically exact solutions of the membrane equilibrium, for a wide range of deflections. 

\section{Second variation spectra} \label{sec:spectra}
We adopt the foregoing variational approximation in this section to determine the second variation spectra corresponding to the in-plane and out-of-plane equilibria.  

\subsection{In-plane spectrum} \label{sec:spectra:inplane}
The spectrum of the membrane in-plane equilibrium is determined by
\begin{equation}
	\left\{ \begin{array}{l}
         		\mathscr{L}_{(\hat{u})}Z=r_{p}\vartheta Z\\
		Z(R_{00})=0,\quad Z(R)=0,\\
	\end{array} \right. 
    	 \label{eqn:eq57}
\end{equation}
where $\vartheta$, $Z$, and $r_{p}$ are the eigenvalue, eigenfunction, and weight function of $\mathscr{L}_{(\hat{u})}$. Non-dimensionalizing Eq.~(\ref{eqn:eq57}) furnishes 
\begin{equation}
	\left\{ \begin{array}{l}
         		\tilde{r}_{p}\frac{\mbox{d}}{\mbox{d}\tilde{r}_{p}}\left( \tilde{r}_{p}\frac{\mbox{d}Z}{\mbox{d}\tilde{r}_{p}}\right)+[(\vartheta^{*}+B)\tilde{r}_{p}^{2}-1]=0\\
		Z(\kappa)=0,\quad Z(1)=0,\\
	\end{array} \right. 
    	 \label{eqn:eq58}
\end{equation}
with $\vartheta^{*}=R^{2}\vartheta/C$ and $B=2(Q_{c}/N_{C})^{4/3}K_{w}/\kappa$. The general solution of Eq.~(\ref{eqn:eq57}) is $Z(\tilde{r}_{p})=C_{1}\mbox{J}_{1}(m\tilde{r}_{p})+C_{2}\mbox{Y}_{1}(m\tilde{r}_{p})$ with $m=\sqrt{B+\vartheta^{*}}$, where $\mbox{J}_{1}$ and $\mbox{Y}_{1}$ are the Bessel functions of first and second kind. A similar eigenvalue problem was derived by \citet{timoshenko2009theory} for the buckling of circular plates under in-plane compressive loads. Imposing the boundary conditions furnishes
\begin{equation}
	Z_{i}(\tilde{r}_{p})=C_{1,i}\left[\mbox{J}_{1}(m_{i}\tilde{r}_{p})-\frac{\mbox{J}_{1}(m_{i}\kappa)}{\mbox{Y}_{1}(m_{i}\kappa)}\mbox{Y}_{1}(m_{i}\tilde{r}_{p})\right],
    	 \label{eqn:eq59}
\end{equation}
\begin{equation}
	\vartheta_{i}^{*}=m_{i}^{2}-B, \quad i=0,1,2,\cdots
    	 \label{eqn:eq60}
\end{equation}
where $m_{i}$ is the $i$th root of
\begin{equation}
	\mbox{J}_{1}(m)\mbox{Y}_{1}(m\kappa)-\mbox{J}_{1}(m\kappa)\mbox{Y}_{1}(m)=0.
    	 \label{eqn:eq61}
\end{equation}

\begin{figure} 
\centering
\includegraphics[scale=0.45]{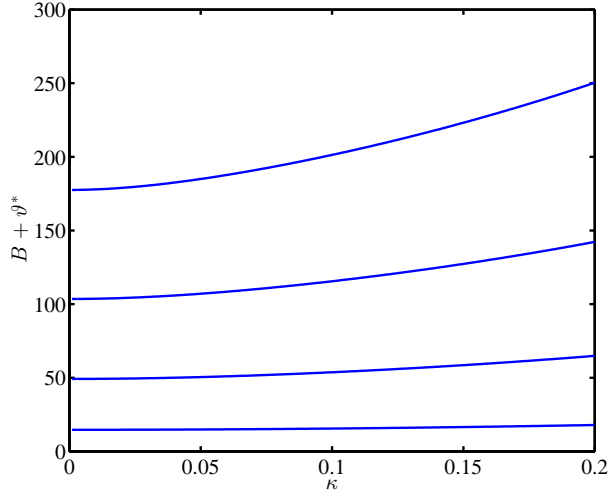}
 \caption{In-plane spectrum $\vartheta_{i}^{*}$ with $i=0,1,2,3$ (upward).}
\label{fig:figure5}
\end{figure}

From Eq.~(\ref{eqn:eq60}), unless $B=m_{0}^{2}$ at a given $\kappa$, $Z_{0}$ corresponding to $\vartheta_{0}^{*}=0$ has only a trivial solution where $\mbox{ker}(\mathscr{L}_{(\hat{u})})=\{ \textbf{0}\}$ and $\dot{\hat{u}}=\textbf{0}$ at $\dot{P}=0$. Here, $\textbf{0}$ is the identically zero function, and the overdot denotes differentiation along equilibrium branches. Therefore, stability loss due to in-plane perturbations is not generally related to pressure turning points. Studying these instabilities, which are responsible for wrinkling in thin elastic membranes \citep{coman2007boundary, davidovitch2011prototypical,pineirua2013capillary}, is beyond the scope of the present work and will not be further elaborated upon.

Figure~\ref{fig:figure5} shows the first four eigenvalues of the in-plane spectrum. Note that $m_{0}^{2}\approx16$ is accurate in the range $0<\kappa<0.2$ to within 12\% of computed values. A positive spectrum is ensured by $B<m_{0}^2$. As water is removed from the elastocapillary system of Fig.~\ref{fig:figure1}, $B$ reaches its maximum value when the membrane touches the plate where $B_{max}=2\Pi^{4}/[K_{w}^{3}\kappa(1-\kappa^2)^{4}]$. To ensure that $\vartheta_{i}^{*}>0$ always holds during drying, we require $B_{max}<m_{0}^{2}$, leading to 

\begin{equation}
	\Pi\lesssim2^{3/4}(1-\kappa^{2}),
    	 \label{eqn:eq62}
\end{equation}
based on the foregoing approximation of $m_{0}$ and the scaling relation $K_{w}^{3}\sim1/\kappa$. Consequently, $N_{C}\gg1/[2^{3/2}(1-\kappa^{2})^{2}]$ guarantees that the in-plane spectrum is positive, and that the membrane profile can be accurately approximated by Eq.~(\ref{eqn:eq55}).

\subsection{Out-of-plane spectrum} \label{sec:spectra:outofplane}
The spectrum of the membrane out-of-plane equilibrium is given by
\begin{equation}
	\left\{ \begin{array}{l}
         		\mathscr{L}_{(\hat{w})}Y=r_{p}\lambda Y\\
		Y'(R_{00})=0,\quad Z(R)=0,\\
	\end{array} \right. 
    	 \label{eqn:eq63}
\end{equation}
where $\lambda$, $Y$, and $r_{p}$ are the eigenvalue, eigenfunction, and weight function of $\mathscr{L}_{(\hat{w})}$. Without attempting to solve Eq.~(\ref{eqn:eq63}), we demonstrate that $\mathscr{L}_{(\hat{w})}$ has a positive spectrum. From the quadratic form
\begin{equation}
	\left< \mathscr{L}_{(\hat{w})}Y_{i},Y_{i}\right>=\lambda_{i}\left< r_{p}Y_{i},Y_{i}\right>=\int_{R_{00}}^{R} r_{p}(\hat{N}_{rr}+C\hat{w}'^{2})Y_{i}'^{2}\mbox{d}r_{p}, \nonumber
\end{equation}
it follows that $\lambda_{i}>0$ because $\hat{N}_{rr}>0$, based on the variational approximation discussed in section~\ref{sec:profile}. Similarly to the in-plane spectrum, $Y_{0}$ corresponding to $\lambda_{0}=0$ has only a trivial solution where $\mbox{ker}(\mathscr{L}_{(\hat{w})})=\{ \textbf{0}\}$ and $\dot{\hat{w}}=\textbf{0}$ at $\dot{P}=0$. 

\subsection{Meniscus spectrum} \label{sec:spectra:meniscus}
When the meniscus is a bridge, the spectrum cannot be determined analytically. Therefore, in this section, we only study the meniscus spectrum for simply connected menisci determined by
\begin{equation}
	\left\{ \begin{array}{l}
         		\mathcal{L}_{(\hat{r})}X=\mu X\\
		y_{0}X'(y_{0})-X(y_{0})=0,\quad X(1)=\mbox{finite},\\
	\end{array} \right. 
    	 \label{eqn:eq64}
\end{equation}
where $\mu$ and $X$ are the eigenvalue\footnote{Not to be confused with the chemical potential in Eqs.~(\ref{eqn:eq1}) and (\ref{eqn:eq2}).} and eigenfunction of $\mathcal{L}_{(\hat{r})}$. These also denote the eigenvalue and eigenfunction of $\mathscr{L}_{(\hat{r})}$ for doubly-connect menisci. Solving Eq.~(\ref{eqn:eq64}) furnishes
\begin{equation}
	X_{i}(y)=C_{1,i}\mbox{P}_{m_{i}}(y),
    	 \label{eqn:eq65}
\end{equation}
\begin{equation}
	\mu_{i}=[(2m_{i}+1)^{2}-9]/4, \quad i=0,1,2,\cdots
    	 \label{eqn:eq66}
\end{equation}
where $\mbox{P}_{m}$ is the real-valued order Legendre function of first kind, and $m_{i}$ is the $i$th root of
\begin{equation}
	y_{0}\mbox{P}'_{m_{i}}(y_{0})-\mbox{P}_{m_{i}}(y_{0})=0.
    	 \label{eqn:eq67}
\end{equation}

\begin{figure} 
\centering
\includegraphics[width=\linewidth]{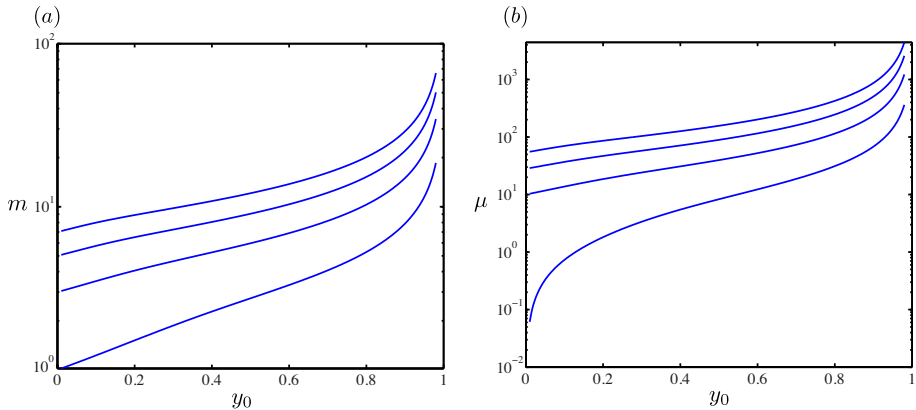}
 \caption{Meniscus spectrum; (a) order of eigenfunctions $m_{i}$ and (b) corresponding eigenvalues $\mu_{i}$ with $i=0,1,2,3$ (upward).}
\label{fig:figure6}
\end{figure}

The boundary condition of Eq.~(\ref{eqn:eq64}) at $y=y_{0}$, resulting from perturbations that can displace the hole edge, is a key feature of this study. It implies that the meniscus contact line at the hole edge exhibits a mixed characteristics of free and pinned contact lines, depending on the bubble size. In the limit $y_{0}\to0$, where the bubble is a hemisphere, the contact line behaves similarly to a pinned contact line. As shown in Fig.~\ref{fig:figure6}, this limit, which corresponds to the pressure turning point ($\dot{P}=0$) of the elastocapillary model in Fig.~\ref{fig:figure1} with simply connected menisci, occurs at the point of stability exchange where $\mu_{0}=0$. Therefore, as expected, this limit coincides with the stability limit of pressure-controlled spherical menisci with a pinned contact line \citep{michael1981meniscus}. Here, the order of the Legendre function takes integer values where $m_{i}=2i+1$, and $\mbox{ker}(\mathcal{L}_{(\hat{r})})\neq\{ \textbf{0}\}$. An important implication is that, unlike the in-plane and out-of-plane spectrum, $\dot{\hat{r}}\neq\textbf{0}$ at $\dot{P}=0$.

\section{Stability along equilibrium branches} \label{sec:branchstability}
In this section, a relation between stability and the slope of equilibrium branches in $v_{l}$ versus $P$ diagrams for constrained and unconstrained problems is established. We assume that the foregoing variational approximation for the membrane equilibrium and, particularly, Eq.~(\ref{eqn:eq62}) always hold, so $\vartheta_{i}>0$ and $\lambda_{i}>0$ for all $i$. As discussed in section \ref{sec:stability}, stability can be determined by the sign of $Q$ in Eqs.~(\ref{eqn:eq43}) and (\ref{eqn:eq44}) with $N,\bar{\xi}_{1},\phi_{1},\psi_{1}\in\mathscr{H}^{0}$ satisfying Eqs.~(\ref{eqn:eq45})-(\ref{eqn:eq48}). Following the Ritz method \citep{gelfand2000calculus}, $Q$ is examined in a countable dense subspace of $\mathscr{H}^{0}$, the existence of which is guaranteed by the separability of $\mathscr{H}^{0}$ \citep{akhiezer1993theory}, because there always exits a function in the dense subspace that is arbitrarily close to any $f\in\mathscr{H}^{0}$. Furthermore, the eigenfunctions of the Sturm-Liouville operators in Eqs.~(\ref{eqn:eq49})-(\ref{eqn:eq52}) form a complete orthogonal basis in $\mathscr{H}^{0}$ \citep{walter1998ordinary}, and, thus, span the respective dense subspace. Hence, $\phi_{1}\in\mbox{span}\{Z_{i}\}$, $\psi_{1}\in\mbox{span}\{Y_{i}\}$, $N\in\mbox{span}\{X_{i}\}$ when the meniscus is a bubble, and $\bar{\xi}_{1}\in\mbox{span}\{X_{i}\}$ when the meniscus is a bridge. 

We unify the representation of quadratic forms in this section by expressing $Q$ in terms of $\bar{\xi}_{1}$ for simply connected and doubly connected menisci. When the meniscus is a bubble, because $\left<\mathscr{L}_{(\hat{r})} \bar{\xi}_{1},\bar{\xi}_{1}\right>=\gamma_{gl}\left<\mathcal{L}_{(\hat{r})} N,N\right>$, the sign of $Q$ is determined by the eigenvalues of $\mathcal{L}_{(\hat{r})}$, no matter how $Q$ is expressed. 

{\parindent0pt 
\textbf{Lemma 1.}  \textit{The necessary condition for an equilibrium branch to be stable in the unconstrained problem is $\mu_{i}\geq0,\vartheta_{i}\geq0,\lambda_{i}\geq0$ for all $i$.}

We assume the contrary is true. For example, let $\mu_{0}<0$ and $\mu_{i}>0$ for $i\geq1$. Therefore, $\bar{\mu}_{0}<0$ . Choosing $\bar{\xi}_{1}=a_{0}\bar{X}_{0}$ with 
\begin{equation}
	a_{0}>\left[ \frac{\left<\bar{\mathscr{L}}_{(\hat{u})} \phi_{1},\phi_{1}\right>_{1}+\left<\bar{\mathscr{L}}_{(\hat{w})} \psi_{1},\psi_{1}\right>_{1}}{|\bar{\mu}_{0}|\left<\bar{X}_{0},\bar{X}_{0}\right>_{1}} \right]^{1/2}
	\nonumber
\end{equation}
results in $Q<0$, which is a contradiction. Here, $\bar{\mu}$ and $\bar{X}$ denote eigenvalues and eigenfunctions of $\bar{\mathscr{L}}_{(\hat{r})}$. $\Box$
}

Note that in lemma 1, Eqs.~(\ref{eqn:eq41})-(\ref{eqn:eq42}) are used to illustrate how the relationship between the spectrum of barred and unbarred operators can be applied to prove stability. Hereafter, Eqs.~(\ref{eqn:eq43})-(\ref{eqn:eq44}) are directly used to determine the sign of $Q$. The following lemma connects stability to the slope of equilibrium branches in the unconstrained problem. 

{\parindent0pt 
\textbf{Lemma 2.}  \textit{The slope of a stable equilibrium branch at any point in the $J$ versus $P$ diagram is non-negative.}

Differentiating Eqs.~(\ref{eqn:eq24})-(\ref{eqn:eq26}) along an equilibrium branch results in
\begin{equation}
	[U''_{(\hat{y})}-PJ''_{(\hat{y})}]\dot{\hat{y}}=\dot{P}J'_{(\hat{y})},\quad \hat{y}=\hat{r},\hat{u},\hat{w},
    	 \label{eqn:eq68}
\end{equation}
furnishing 
\begin{multline}
	\left<[U''_{(\hat{r})}-PJ''_{(\hat{r})}]\dot{\hat{r}},\dot{\hat{r}} \right>+\left<[U''_{(\hat{u})}-PJ''_{(\hat{u})}]\dot{\hat{u}},\dot{\hat{u}} \right>+\left<[U''_{(\hat{w})}-PJ''_{(\hat{w})}]\dot{\hat{w}},\dot{\hat{w}} \right> \\
	=\dot{P}\left[\left<J'_{(\hat{r})},\dot{\hat{r}} \right>+\left<J'_{(\hat{u})},\dot{\hat{u}} \right>+\left<J'_{(\hat{w})},\dot{\hat{w}} \right>\right].
	\nonumber
\end{multline}
Using Eq.~(\ref{eqn:eqb2}),
\begin{multline}
	\dot{J}=\left<J'_{(\hat{r})},\dot{\hat{r}} \right>+\left[\dot{\hat{r}}K_{r'}+\dot{\hat{z}}K \right]_{\hat{z}_{0}}^{\hat{h}}+\left<J'_{(\hat{u})},\dot{\hat{u}} \right>+
	\left[\dot{\hat{u}}M_{u'}\right]_{R_{00}}^{R}\\
	+\left<J'_{(\hat{w})},\dot{\hat{w}} \right>+\left[\dot{\hat{w}}M_{w'}\right]_{R_{00}}^{R}+\hat{R}_{0}\dot{\hat{R}}_{0}\hat{h}+\frac{\hat{R}_{0}^{2}\dot{\hat{h}}}{2}.
	\nonumber
\end{multline}
Substituting for $K$ and $M$ from Eqs.~(\ref{eqn:eq9}) and (\ref{eqn:eq11}), all the boundary terms cancel each other, leading to
\begin{equation}
	\dot{J}=\left<J'_{(\hat{r})},\dot{\hat{r}} \right>+\left<J'_{(\hat{u})},\dot{\hat{u}} \right>+\left<J'_{(\hat{w})},\dot{\hat{w}} \right>
	\nonumber
\end{equation}
and, consequently,
\begin{multline}
	\left<[U''_{(\hat{r})}-PJ''_{(\hat{r})}]\dot{\hat{r}},\dot{\hat{r}} \right>+\left<[U''_{(\hat{u})}-PJ''_{(\hat{u})}]\dot{\hat{u}},\dot{\hat{u}} \right>+\left<[U''_{(\hat{w})}-PJ''_{(\hat{w})}]\dot{\hat{w}},\dot{\hat{w}} \right> \\
	=\dot{P}\dot{J}=\dot{P}^{2}\frac{\mbox{d}J}{\mbox{d}P}.
	\nonumber
\end{multline}
Since the branch is stable, $\mu_{i}\geq0,\vartheta_{i}\geq0,\lambda_{i}\geq0$ according to Lemma 1. Thus, the left-hand side is non-negative. $\Box$
}

Similarly, the following lemma connects stability to the slope of equilibrium branches in the constrained problem. 

{\parindent0pt 
\textbf{Lemma 3.}  \textit{Suppose that, on a segment of an equilibrium branch, $\mathscr{L}_{(\hat{r})}$ is non-singular and has precisely one negative eigenvalue. Then, the segment is stable if and only if the slope at any point in the $J$ versus $P$ diagram is negative.}

We prove the lemma for doubly connected menisci. Consider the following perturbation decompositions
\begin{equation}
	\bar{\xi}_{1}=v_{r}+\alpha\eta_{r}, \quad\phi_{1}=v_{u}+\alpha\eta_{u},\quad\psi_{1}=v_{w}+\alpha\eta_{w},
	\nonumber
\end{equation}
where
\begin{equation}
	\mathscr{L}_{(\hat{y})}\eta_{y}=J'_{(\hat{y})}, \quad y=r,u,w.
    	 \label{eqn:eq69}
\end{equation}
Because $\mathscr{L}_{(\hat{y})}$ are all non-singular, $\mbox{ker}(\mathscr{L}_{(\hat{y})})=\{\textbf{0}\}$ and $J'_{(\hat{y})}\in\mbox{ker}(\mathscr{L}_{(\hat{y})})^{\bot}$. Therefore, $\eta_{y}$ always have a solution because $\mathscr{L}_{(\hat{y})}$ are Fredholm operators. From the volume constraint in Eq.~(\ref{eqn:eq53}),
\begin{multline}
	\left<\mathscr{L}_{(\hat{r})} \eta_{r},v_{r}\right>+\left<\mathscr{L}_{(\hat{u})} \eta_{u},v_{u}\right>+\left<\mathscr{L}_{(\hat{w})} \eta_{w},v_{w}\right>\\
	=-\alpha\left[\left<\mathscr{L}_{(\hat{r})} \eta_{r},\eta_{r}\right>+\left<\mathscr{L}_{(\hat{u})} \eta_{u},\eta_{u}\right>+\left<\mathscr{L}_{(\hat{w})} \eta_{w},\eta_{w}\right> \right],
	\nonumber
\end{multline}
furnishing
\begin{multline}
	Q=\left<\mathscr{L}_{(\hat{r})} v_{r},v_{r}\right>+\left<\mathscr{L}_{(\hat{u})} v_{u},v_{u}\right>+\left<\mathscr{L}_{(\hat{w})} v_{w},v_{w}\right>\\
	-\alpha^{2}\left[\left<\mathscr{L}_{(\hat{r})} \eta_{r},\eta_{r}\right>+\left<\mathscr{L}_{(\hat{u})} \eta_{u},\eta_{u}\right>+\left<\mathscr{L}_{(\hat{w})} \eta_{w},\eta_{w}\right> \right].
	\nonumber
\end{multline}
Note that $\left<\mathscr{L}_{(\hat{u})} \eta_{u},\eta_{u}\right>,\left<\mathscr{L}_{(\hat{w})} \eta_{w},\eta_{w}\right>,\left<\mathscr{L}_{(\hat{u})} v_{u},v_{u}\right>,\left<\mathscr{L}_{(\hat{w})} v_{w},v_{w}\right> >0$ because the in-plane and out-of-plane spectrum are positive. We first show that the necessary condition for $Q>0$ is
\begin{equation}
	\left<\mathscr{L}_{(\hat{r})} \eta_{r},\eta_{r}\right>+\left<\mathscr{L}_{(\hat{u})} \eta_{u},\eta_{u}\right>+\left<\mathscr{L}_{(\hat{w})} \eta_{w},\eta_{w}\right><0.
    	 \label{eqn:eq70}
\end{equation}
We assume the contrary holds. Choosing $v_{r}=a_{0}X_{0}$ with 
\begin{equation}
	a_{0}>\left[ \frac{\left<\mathscr{L}_{(\hat{u})} v_{u},v_{u}\right>+\left<\mathscr{L}_{(\hat{w})} v_{w},v_{w}\right>}{|\mu_{0}|\left<X_{0},X_{0}\right>} \right]^{1/2}
	\nonumber
\end{equation}
results in $Q<0$, which is a contradiction. Therefore, it is always possible to construct perturbations that lead to instability if Eq.~(\ref{eqn:eq70}) does not hold. Here, $\mu_{0}$ and $X_{0}$ are the negative eigenvalue and corresponding eigenfunction of $\mathscr{L}_{(\hat{r})}$. Next, we show that Eq.~(\ref{eqn:eq70}) is sufficient for $Q>0$. Since perturbations are selected from a countable dense space, $\eta_{r}$ can be written $\eta_{r}=b_{0}X_{0}+\sum_{i=1}^{\infty}b_{i}X_{i}$ such that 
\begin{equation}
	b_{0}>\left[ \frac{\sum_{i=1}^{\infty}b_{i}^2\mu_{i}\left< X_{i},X_{i}\right>+\left<\mathscr{L}_{(\hat{u})} \eta_{u},\eta_{u}\right>+\left<\mathscr{L}_{(\hat{w})} \eta_{w},\eta_{w}\right>}{|\mu_{0}|\left<X_{0},X_{0}\right>} \right]^{1/2}
	\nonumber
\end{equation}
for Eq.~(\ref{eqn:eq70}) to hold. Moreover, any arbitrary meniscus perturbation can be written $\bar{\xi}_{1}=a_{0}X_{0}+\sum_{i=1}^{\infty}a_{i}X_{i}$. Choosing $\alpha=a_{0}/b_{0}$ leads to $v_{r}=\sum_{i=1}^{\infty}c_{i}X_{i}$, implying that $\left<\mathscr{L}_{(\hat{r})} v_{r},v_{r}\right>>0$, and, consequently, $Q>0$. Hence, all meniscus perturbations, including those with $\left<\bar{\xi}_{1},X_{0}\right>\neq0$, can be decomposed into $\eta_{r}$ and $v_{r}$, leading to a strictly positive second variation, provided Eq.~(\ref{eqn:eq70}) holds.

Substituting Eq.~(\ref{eqn:eq69}) into Eq.~(\ref{eqn:eq68}) furnishes
\begin{equation}
	\eta_{y}=\dot{\hat{y}}/\dot{P}, \quad y=r,u,w,
	\nonumber
\end{equation}
giving
\begin{multline}
	\left<\mathscr{L}_{(\hat{r})} \eta_{r},\eta_{r}\right>+\left<\mathscr{L}_{(\hat{u})} \eta_{u},\eta_{u}\right>+\left<\mathscr{L}_{(\hat{w})} \eta_{w},\eta_{w}\right>\\
	=\frac{1}{\dot{P}}\left[\left<J'_{(\hat{r})},\dot{\hat{r}} \right>+\left<J'_{(\hat{u})},\dot{\hat{u}} \right>+\left<J'_{(\hat{w})},\dot{\hat{w}} \right>\right]=\frac{\dot{J}}{\dot{P}}=\frac{\mbox{d}J}{\mbox{d}P},
	\nonumber
\end{multline}
which completes the proof. $\Box$
}

\begin{figure} 
\centering
\includegraphics[width=\linewidth]{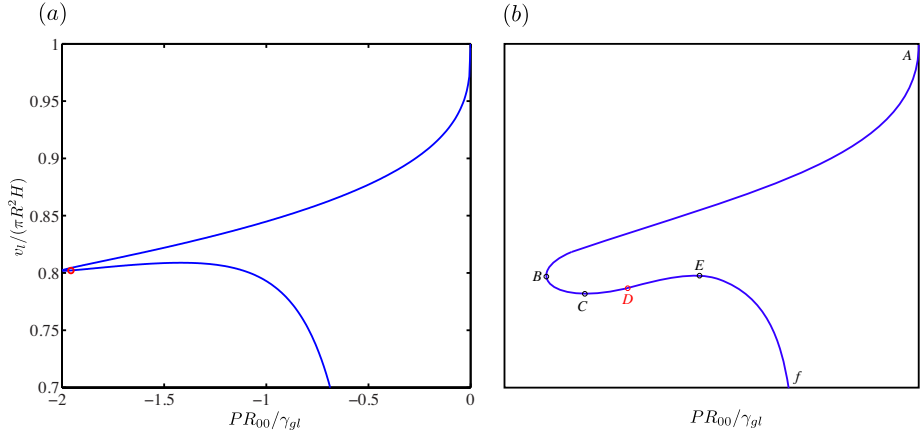}
 \caption{Equilibrium branch of the elastocapillary model in Fig.~\ref{fig:figure1} for simply connected menisci with $\kappa=0.1,\nu=0.3,N_{C}=15000,\Pi=0.2$; (a) numerical computation and (b) schematic representation of pressure and volume turning points.}
\label{fig:figure7}
\end{figure}

\begin{figure} [t]
\centering
\includegraphics[width=\linewidth]{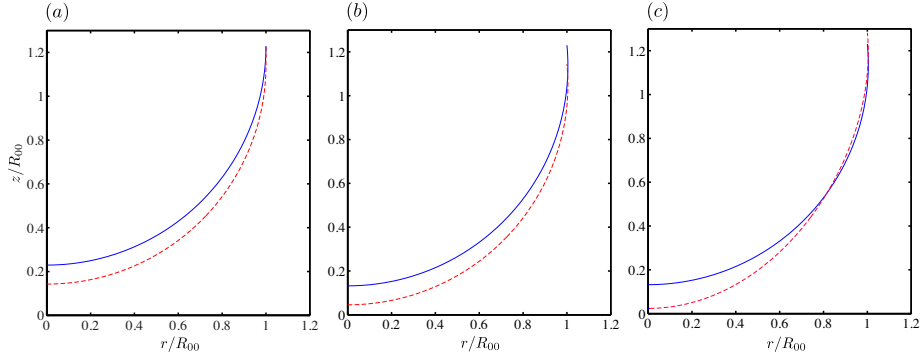}
 \caption{Meniscus equilibrium (solid) and perturbed (dashed) states at volume and pressure turning points in Fig.~\ref{fig:figure7}; The most dangerous perturbation normalized by $\left<N,N\right>=R_{00}^2$ at (a) the pressure turning point $B$ where $N(y)=\sqrt{3}R_{00}\mbox{P}_{1}(y)$ and (b) the volume turning point $C$ where $N(y)=\sqrt{1-y_{0}^2}/\left<\mbox{P}_{m_{0}},\mbox{P}_{m_{0}}\right>R_{00}\mbox{P}_{m_{0}}(y)$. (c) A safe perturbation $N(y)=C_0X_{0}(y)+C_1X_{1}(y)$ at $C$ normalized by $\left<N,N\right>+\left<\psi_{1},\psi_{1}\right>=R_{00}^2$ where $\phi_{1}(\tilde{r}_{p})=0$ and $\psi_{1}(\tilde{r}_{p})=a_0+a_1\tilde{r}_{p}+a_2\tilde{r}_{p}^{2}$ such that Eqs.~(\ref{eqn:eq16})-(\ref{eqn:eq20}) and (\ref{eqn:eq53}) are satisfied, leading to $Q>0$.}
\label{fig:figure8}
\end{figure}

The proof of lemma 3 for simply connected menisci similarly proceeds using the decomposition $N=v_{rn}+\alpha\eta_{rn}$ and accounting for the relation between $N$ and $\bar{\xi}_{1}$ given by Eq.~(\ref{eqn:eqa4}). Figure~\ref{fig:figure7} shows how lemmas 1-3 can be applied to determine stability from the shape of an equilibrium branch in the unconstrained and constrained problems. In the unconstrained problem, where $P$ is the control parameter, only the segments $AB$ and $CE$ can be stable according to lemma 2. The stability of the segment $AB$, excluding $B$, is deduced from the system configuration at $A$, corresponding to the fully saturated state. Here, the meniscus and membrane are planar, and the system is evidently stable, implying that $\mu_{i},\vartheta_{i},\lambda_{i}>0$ for all $i$. Assuming that $\mu_{0}$ varies continuously along the equilibrium branch, stability is lost at $B$, and the entire segment $BCDEf$ is unstable. The segment $AB$ is also stable in the constrained problem, where $v_{l}$ is the control parameter. Moreover, the stability of $BC$, along which $\mathcal{L}_{(\hat{r})}$ has one negative eigenvalue, is deduced from lemma 3. Beyond the volume turning point $C$, the entire segment $CDEf$ is unstable with respect to constant-volume perturbations.  

Determining the stability of equilibrium branches is essential for predicting the dry-state conformation of the present model. This is illustrated by an example in Fig.~\ref{fig:figure7}. Here, the point $D$ corresponds to the state where the bubble is tangent to the plate. Further decrease in $v_{l}$ forces the bubble to bridge the membrane and plate, which is a necessary step for the elastocapillary system of Fig.~\ref{fig:figure1} to collapse. However, when drying from the fully saturated state at $A$, collapse does not occur for the given parameters in Fig.~\ref{fig:figure7}, because the system loses stability at $C$ before the bubble can bridge the membrane and plate. 

Note that spherical menisci with a pinned contact line are always stable to constant-volume perturbations \citep{myshkis1987low}. Furthermore, as discussed in section~\ref{sec:spectra}, the membrane is stable for all deflections, provided Eq.~(\ref{eqn:eq62}) is satisfied. Therefore, the meniscus and membrane are individually stable along the entire branch $ABCDEf$. However, the elastocapillary system as a whole subject to $v_{l}=\mbox{const.}$ is unstable along $CDEf$, revealing an intimate connection between stability and the coupling of elastic and capillary forces. This manifests in the boundary shared by the elastic and capillary part where the meniscus and membrane interact through boundary displacing perturbations. Moreover, the nature of instabilities are influenced by the control parameter and how the meniscus and membrane interact with each other, as demonstrated in Fig.~\ref{fig:figure8}. 

\section{Concluding remarks} \label{sec:conclusion}

We have developed an elastocapillary model to study drying-induced structural failures, such as those arising from stiction in microelectromechanical systems. The model comprises an elastic membrane and a meniscus, deformed by the same pressure differential, interacting through a shared boundary. The existence of a stable equilibrium branch from the fully saturated to collapsed state is an essential precursor for structural failures. We examined the model stability and equilibrium using variational and spectral methods. Stability was related to the slope of equilibrium branches in the liquid content versus pressure diagram for the constrained and unconstrained problems. A variational approximation, complemented by scaling analysis, was derived, furnishing closed-form expressions for membrane equilibria. This approximation leads to a positive out-of-plane spectrum. For a given geometry, there is a critical elstocapillary number above (below) which the in-plane spectrum is positive (has a negative eigenvalue). These in-plane instabilities are a common cause of wrinkling in thin membranes. Thus, except for thin membranes, only meniscus perturbations can be dangerous for the elstocapillary system. This paper extends the work of \citet{maddocks1987stability} to elastocapillary systems that are subjected to boundary displacing perturbations, revealing a close connection between stability and the coupling of elastic and capillary forces. We demonstrated that the stability of the meniscus and membrane alone does not imply that the elastocapillary system as a whole is stable; the destabilizing effect of the membrane and meniscus interacting through their shared boundary must also be accounted for. Moreover, our results support a general concept in catastrophe theory that stability exchanges occur at turning points with respect to the control parameter, thereby, reducing costly stability computations to search for folds on equilibrium branches.

\numberwithin{equation}{section}
\begin{appendices}
\section{} \label{sec:appendixa}
Expressing perturbations in $\bar{\xi}_{1}$ is problematic for axisymmetric simply connected menisci because $\bar{\xi}_{1}\not\in L^{2}$. Assuming that perturbed states are also axisymmetric (\ie, $\mbox{d}z/\mbox{d}r=0$ at $\hat{z}=\hat{\ell}$), we have
\begin{equation}
	\frac{\mbox{d}z}{\mbox{d}r}=\frac{1+\eta'_{1}\varepsilon+\cdots}{\hat{r}'+\xi'_{1}\varepsilon+\cdots}=\frac{1}{\hat{r}'}-\frac{\bar{\xi}'_{1}+\eta_{1}\hat{r}''}{\hat{r}'^{2}}\varepsilon+\cdots=0 \quad \mbox{at} \quad \hat{z}=\hat{\ell}
	\label{eqn:eqa1}
\end{equation}
in view of Eqs.~(\ref{eqn:eq12}) and (\ref{eqn:eq13}). Since $\hat{r}'(\hat{\ell})\to\infty$ and $\xi_{1}(\hat{\ell}),\eta_{1}(\hat{\ell}),\hat{r}''(\hat{\ell})=\mbox{finite}$, it follows that $|\bar{\xi}'_{1}(\hat{\ell})|\sim O(\hat{r}')\to\infty$ satisfies the axial symmetry condition. Moreover, From Eq.~(\ref{eqn:eq21}), $|\bar{\xi}_{1}(\hat{\ell})|\to\infty$. Therefore, $\bar{\xi}$ is unbounded, and, thus, unsuitable for representing perturbations when the meniscus is a bubble. Here, representing functional variations with respect to the normal variations of menisci resolves the issue. 

When the meniscus is a bubble (sphere), it is convenient to express variables as functions of the meridian-curve arclength $\hat{s}$ 
or the polar angle $\theta=\hat{s}/R_{s}$ (see Fig.~\ref{fig:figure2}). The displacement vector from the meniscus equilibrium states to its perturbed states is written 
\begin{equation}
	\frac{\delta\textbf{x}}{\varepsilon}=\xi_{1}\hat{\textbf{r}}+\eta_{1}\hat{\textbf{z}}.
	\label{eqn:eqa2}
\end{equation}
Given $\hat{\textbf{n}}=\sin \theta \hat{\textbf{r}}-\cos \theta \hat{\textbf{z}}$, the normal variation
\begin{equation}
	N=\hat{\textbf{n}}\cdot\frac{\delta\textbf{x}}{\varepsilon}=\xi_{1}\sin\theta-\eta_{1}\cos\theta
	\label{eqn:eqa3}
\end{equation}
is obtained, furnishing 
\begin{equation}
	\bar{\xi}_{1}=\frac{N}{\sin\theta},
	\label{eqn:eqa4}
\end{equation}
resulting in $\mbox{d}N/\mbox{d}\hat{s}=-\mbox{d}\eta_{1}/\mbox{d}\hat{s}$ at $\hat{z}=\hat{\ell}$. Moreover, $\eta'_{1}\to\xi'_{1}/\hat{r}'$ and $\mbox{d}\eta_{1}/\mbox{d}\hat{s}\to\xi'_{1}/\hat{r}'\sqrt{1+\hat{r}'^{2}}$ as $\hat{z}\to\hat{\ell}$, giving $\mbox{d}\eta_{1}/\mbox{d}\hat{s}(\hat{\ell})\to0$ even if $|\xi'_{1}|\sim O(\hat{r}')$. Therefore, $\mbox{d}N/\mbox{d}\hat{s}=0$ and $N=\mbox{finite}$ at $\hat{z}=\hat{\ell}$, implying $N\in L^{2}$.

The following formulas are useful for representing the meridian curve when the meniscus is a bubble:
 \begin{equation}
 	\hat{r}=\sqrt{R_{s}^{2}-(z_{c}-\hat{z})^{2}}=R_{s}\sin\theta,
	\label{eqn:eqa5}
\end{equation}
 \begin{equation}
 	\hat{r}'=\frac{z_{c}-\hat{z}}{\hat{r}}=\cot \theta,
	\label{eqn:eqa6}
\end{equation}
 \begin{equation}
 	\hat{r}''=-\frac{1+\hat{r}'^{2}}{\hat{r}}=-\frac{1}{R_{s}\sin^{3}\theta},
	\label{eqn:eqa7}
\end{equation}
furnishing 
 \begin{equation}
 	\mathcal{P}^{(\hat{r})}=R_{s}\gamma_{gl}\sin^{4}\theta,\quad \mathcal{Q}^{(\hat{r})}=-\frac{\gamma_{gl}}{R_{s}},\quad \mathcal{A}=\frac{\gamma_{gl}\sin^{2}\theta}{\cos\theta},
	\label{eqn:eqa8}
\end{equation}
where $z_{c}$ and $R_{s}$ are the $z$-coordinate at the center and radius of the sphere.  

\section{} \label{sec:appendixb}
Consider the functional 
 \begin{equation}
 	J[y]=\int_{x_a}^{x_b} F(x,y,y')\mbox{d}x, \quad J:L^{2}\rightarrow\mathbb{R},\quad y:[x_a,x_b]\rightarrow\mathbb{R},
	\label{eqn:eqb1}
\end{equation}
of continuously differentiable functions $y$ defined on a variable domain where the branches of stationary points are parametrized with $t$, and the stationary points are represented by $\hat{y}=\hat{y}(\hat{x},t)$. Then, differentiating the functional and its first Fr\'echet derivative along a branch furnishes 
 \begin{equation}
 	\dot{J}=\left< J'_{(\hat{y})},\dot{\hat{y}} \right>+\left[\dot{\hat{y}}F_{y'}+\dot{\hat{x}}F \right]_{\hat{x}_a}^{\hat{x}_b},
	\label{eqn:eqb2}
\end{equation}
 \begin{equation}
 	\dot{J}'_{(\hat{y})}=J''_{(\hat{y})}[\dot{\hat{y}}],
	\label{eqn:eqb3}
\end{equation}
where
\begin{equation}
 	J''_{(\hat{y})}[\varphi]=-\frac{\mbox{d}}{\mbox{d}\hat{x}}\left(F_{y'y'} \frac{\mbox{d}\varphi}{\mbox{d}\hat{x}}\right)+\left(F_{yy} -\frac{\mbox{d}}{\mbox{d}\hat{x}}F_{yy'} \right)\varphi,\quad J''_{(\hat{y})}:L^{2}\rightarrow L^{2}.
	\label{eqn:eqb4}
\end{equation}
 
\end{appendices}

\section*{Acknowledgements} \label{sec:acknowledgements}

Support from the NSERC Innovative Green Wood Fibre Products Network and the McGill University Faculty of Engineering is gratefully acknowledged.

\bibliography{mybib}
\bibliographystyle{jfm}

\end{document}